\documentclass[entropy,article,accept,moreauthors,10pt,a4paper]{mdpi}
\firstpage{1}
    \pubvolume{00}
    \issuenum{0}
\articlenumber{00}
\pubyear{0000}
\copyrightyear{0000}
\dateaccepted{00 XX 0000}
\datepublished{00 XX 0000}
\hreflink{https://doi.org/10.3390/e00000000} % If needed use \linebreak
\newcommand\beq{\begin{equation}}
\newcommand\eeq{\end{equation}}
\newcommand\beqa{\begin{eqnarray}}
\newcommand\eeqa{\end{eqnarray}}
\newcommand{\nn}{\nonumber\\}
\def\bal#1\eal{\begin{align}#1\end{align}}
    \newcommand{\Tb}{T_{\text{b}}}
    \newcommand{\TA}{T_{\text{A}}}
    \newcommand{\TB}{T_{\text{B}}}
    \newcommand{\Th}{T_{\text{hot}}}
    \newcommand{\Tc}{T_{\text{cold}}}
    \newcommand{\Tw}{T_{\text{warm}}}

    \newcommand{\tw}{t_{\text{w}}}
    \newcommand{\twA}{t_{\text{w}}^{\text{A}}}
    \newcommand{\twB}{t_{\text{w}}^{\text{B}}}
    \newcommand{\twI}{t_{\text{w}}^{\text{I}}}
    \newcommand{\twII}{t_{\text{w}}^{\text{II}}}
    \newcommand{\ttw}{\widetilde{t}_{\text{w}}}
    \newcommand{\bartw}{\overline{t}_{\text{w}}}
    \newcommand{\uw}{u_{\text{w}}}
    \newcommand{\BI}{B1}
    \newcommand{\BII}{B2}
    \newcommand{\BIII}{B3}
    \newcommand{\BIV}{B4}
    \newcommand{\tx}{t_{\times}}
       \newcommand{\ww}{\omega}
\newcommand{\EE}{\mathcal{E}}

\newcommand{\ee}{\text{e}}
\newcommand{\wT}{\widetilde{T}}
\newcommand{\wTb}{\widetilde{T}_{\text{b}}}
\newcommand{\wEE}{\widetilde{\mathcal{E}}}
\newcommand{\TI}{T_{\text{I}}}
\newcommand{\TII}{T_{\text{II}}}
\newcommand{\TIII}{T_{\text{III}}}
\newcommand{\TIA}{\TI^{\text{A}}}
\newcommand{\TIB}{\TI^{\text{B}}}
\newcommand{\TIIA}{\TII^{\text{A}}}
\newcommand{\TIIB}{\TII^{\text{B}}}
\newcommand{\TIIIB}{\TIII^{\text{B}}}
\newcommand{\alAB}{\alpha}
\newcommand{\beAB}{\beta}
\newcommand{\talAB}{\widetilde{\alpha}}
\newcommand{\aAB}{a_{\text{I}}}
\newcommand{\bAB}{a_{\text{II}}}
\newcommand{\tDelta}{\widetilde{\Delta}}
\newcommand{\tD}{\widetilde{D}}

\Title{The role of preparation protocols in the Mpemba effect under time-delayed Newton cooling}

\TitleCitation{The role of preparation protocols in the Mpemba effect under time-delayed Newton cooling}

\Author{{Andr\'es Santos} $^{1,2,}$*\orcidC{}}

\AuthorNames{Andr\'es Santos}

\AuthorCitation{Santos, A.}

\address{%
$^{1}$ \quad Departamento de F\'isica, Universidad de Extremadura, {E-06006} Badajoz, Spain \\
$^{2}$ \quad Instituto de Computaci\'on Cient\'ifica Avanzada (ICCAEx), Universidad de
Extremadura, \mbox{E-06006 Badajoz, Spain}}

\corres{Correspondence: andres@unex.es}

\abstract{
The Mpemba effect depends on both the relaxation dynamics and the protocol used to prepare the initial states. Here, this dependence is investigated using the time-delayed Newton cooling law, comparing protocols in which two samples are exposed to piecewise-constant bath temperatures before being placed in a common final bath. For protocols with at most two quenches and a common waiting time, exact conditions for the occurrence of the effect are derived. They rule out the Pontus protocol within this model and show that the synchronous two-reservoir protocol maximizes the effect magnitude within the class considered. The robustness of this optimum is then examined by allowing asynchronous quenches or an additional quench for one sample. Numerical optimization of these extensions identifies the synchronous two-reservoir protocol as the maximizing limit in the cases studied. This evidence motivates a conjecture that the same protocol remains optimal for more general bath histories, although that broader statement is not proved.
}


\begin{document}

%%%%%%%%%%%%%%%%%%%%%%%%%%%%%%%%%%%%%%%%%%
%\setcounter{section}{-1} %% Remove this when starting to work on the template.

\renewcommand{\topfraction}{0.9}       % max share of a page for top floats (default 0.7)
\renewcommand{\bottomfraction}{0.9}    % same for bottom floats (default 0.3)
\renewcommand{\textfraction}{0.07}     % min share of a page for text (default 0.2)
\renewcommand{\floatpagefraction}{0.8} % a float-only page must be at least 80% full (default 0.5)
\setcounter{topnumber}{3}
\setcounter{bottomnumber}{2}
\setcounter{totalnumber}{5}

\section{Introduction}
\label{sec1}

The Mpemba effect (ME) refers to the counterintuitive possibility that, under  suitable conditions, a system prepared at a higher temperature can reach a prescribed lower temperature sooner than one prepared at a lower temperature. First brought to broad attention by Mpemba and Osborne's experiments on water~\cite{MO69}, the effect has since prompted debate about its mechanisms and reproducibility in that setting~\cite{BL16,BH20,SHZMW23}. It is now studied more broadly as an anomalous feature of nonequilibrium relaxation, with Mpemba-like behavior reported or predicted in systems ranging from granular gases and colloids to stochastic and quantum models~\cite{LVPS17,LR17,TLLVPS19,KB20,BKC21,CLL21,CTH23,D23,ACM25,NE25,NEDG25,TBLRV26,L26,L26b}. In these settings, the effect generally depends on how the initial states are prepared and on the subsequent relaxation dynamics. The preparation protocol is therefore part of the physical question, rather than a detail that can be left implicit.

The role of preparation is especially clear in simple cooling models. Under Newton's law of cooling with a constant bath temperature, two samples with different initial temperatures approach the bath exponentially, and their temperature difference retains its sign. This memoryless description cannot produce a Mpemba crossover. A minimal way to introduce memory is to let the instantaneous cooling rate depend on the sample temperature at an earlier time. The resulting time-delayed Newton's law provides a tractable phenomenological model for examining how memory and bath history affect thermal relaxation~\cite{S24,S25,S26b}.

In earlier work, the time-delayed cooling law was used to study the Mpemba and Kovacs effects under a two-reservoir protocol~\cite{S24}. That analysis identified the conditions for a Mpemba crossover in terms of the delay time and a common waiting time, and showed that the same model also produces an anomalous Kovacs response. The subsequent study considered finite-rate thermal quenches, examining how a nonzero quench duration modifies both effects~\cite{S25}.
The Descartes protocol was then studied as a three-reservoir preparation in which the two samples undergo single quenches at different times~\cite{S26b}; its additional temperature parameter allowed the effect to be optimized and compared directly with the two-reservoir result. These studies establish that predictions within the delayed Newton model can depend sensitively on the chosen protocol.

The present paper compares several preparations within a common class of bath histories involving at most two quenches and a shared waiting time. Alongside the two-reservoir and Descartes protocols, the comparison includes the Pontus protocol~\cite{NE25,NEDG25,L26b} and two further schedules, denoted WHC and CWC. Exact solutions of the delayed cooling law yield necessary and sufficient conditions for an ME in this class. These conditions exclude the Pontus protocol, while allowing the effect for the two-reservoir and Descartes protocols. The effect magnitude is then optimized over the admissible reservoir temperatures and waiting time. Within the stated class, the synchronous two-reservoir protocol is shown to give the largest magnitude.

The final part of the paper asks whether this optimum persists when the restrictions defining that class are relaxed. First, the two-reservoir protocol is considered with different waiting times for the two samples. Next, four three-quench schedules for the initially cooler sample  are examined. Numerical optimization in these cases finds the largest effect when the schedules reduce to the synchronous two-reservoir protocol. This evidence motivates, but does not prove, the conjecture that the same protocol is optimal over a broader class of preparation histories.

The paper is organized as follows. Section~\ref{sec2} defines the preparation protocols considered, and Sec.~\ref{sec3} introduces the time-delayed Newton cooling law. The conditions for the ME are derived in Sec.~\ref{sec4}, followed by the optimization of its magnitude within the common-wait, two-quench class in Sec.~\ref{sec5}. Section~\ref{sec6} examines asynchronous quenches and three-quench schedules, and discusses the conjecture for more general preparation histories. Conclusions are presented in Sec.~\ref{sec7} by stating the scope of the analytical result, the numerical evidence, and the conjecture separately. The Appendix gives the solution of the model for a generic time-dependent bath.

\section{Protocols}
\label{sec2}

In the ME, two samples, A and B, are prepared according to different protocols for $t<0$, with
\beq
\label{1}
\TA(0)>\TB(0).
\eeq
For $t>0$, both samples are placed in contact with the same reservoir at temperature $\Tc<\TB(0)$. The ME occurs if the initially hotter sample A becomes colder than sample B at a later time. We restrict attention to cases in which neither sample overshoots the target temperature, so that $T_{\text{X}}(t)>\Tc$ for $\text{X}=\mathrm{A},\mathrm{B}$ and $t>0$, and the sample temperatures cross only once. Thus, more complex scenarios \cite{THS21,MS22b} are not considered. Denoting the crossover time by $\tx$, this condition reads
\beq
\label{2}
\TA(t)>\TB(t) \quad (0\leq t<\tx), \quad
\TA(t)<\TB(t) \quad (t>\tx).
\eeq

\begin{figure}
\begin{adjustwidth}{-\extralength}{0cm}
\centering
\includegraphics[width=0.32\linewidth]{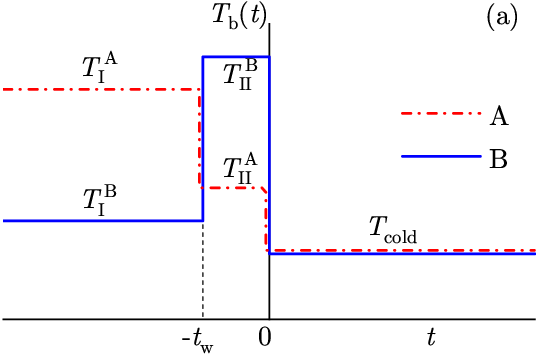}\hfill
\includegraphics[width=0.32\linewidth]{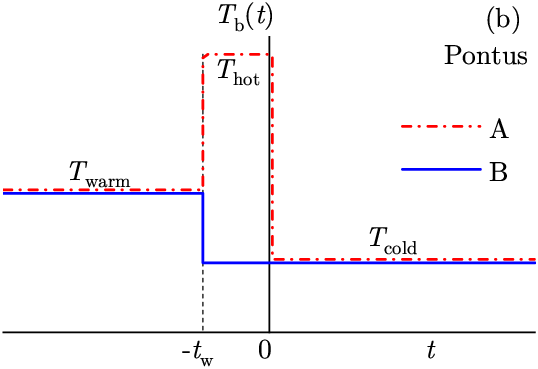}\hfill
\includegraphics[width=0.32\linewidth]{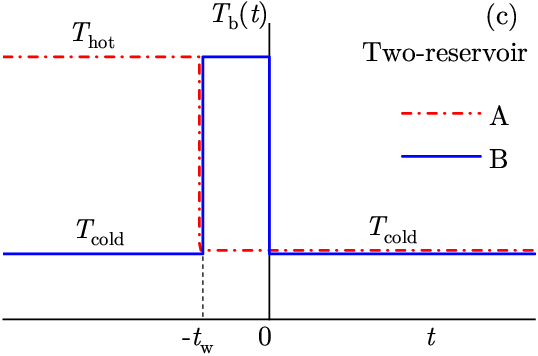}\\[0.4cm]
\includegraphics[width=0.32\linewidth]{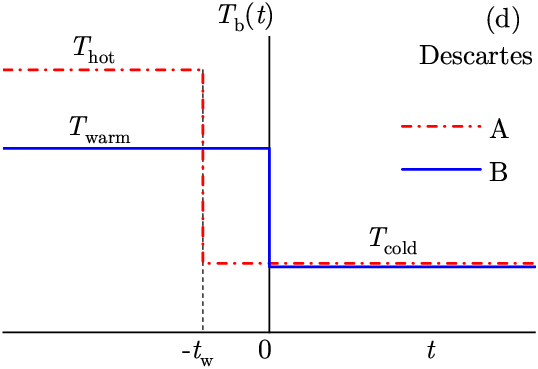}\hfill
\includegraphics[width=0.32\linewidth]{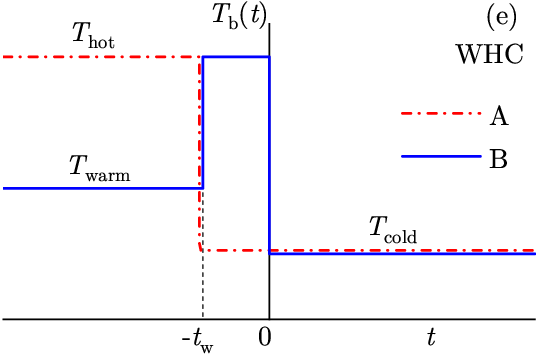}\hfill
\includegraphics[width=0.32\linewidth]{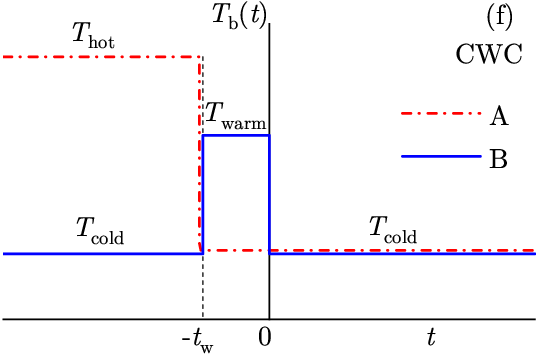}
\end{adjustwidth}
\vspace{-3pt}
\caption{Schematic reservoir-temperature schedules for samples A (red dash-dotted lines) and B (blue solid lines). In every panel, both samples are placed in contact with the reservoir at temperature $\Tc$ for $t>0$. (a) Class of protocols with four reservoir temperatures $\{\TIA,\TIIA,\TIB,\TIIB\}$ for $t<0$, and a common waiting time $\tw$. (b) Pontus protocol~\cite{NE25,NEDG25}: both samples are initially at $\Tw$; sample A is then exposed to $\Th$ before the final quench to $\Tc$, while sample B is quenched directly to $\Tc$ at $t=-\tw$. (c) Two-reservoir protocol~\cite{S24,S25}: sample A is quenched from $\Th$ to $\Tc$ at $t=-\tw$; sample B is quenched from $\Tc$ to $\Th$ at $t=-\tw$ and back to $\Tc$ at $t=0$. (d) Descartes protocol~\cite{S26b}: sample A is quenched from $\Th$ to $\Tc$ at $t=-\tw$, while sample B is quenched from $\Tw$ to $\Tc$ at $t=0$. (e) WHC protocol: sample A is quenched from $\Th$ to $\Tc$ at $t=-\tw$, while sample B follows $\Tw\to\Th\to\Tc$. (f) CWC protocol: sample A is quenched from $\Th$ to $\Tc$ at $t=-\tw$, while sample B follows $\Tc\to\Tw\to\Tc$.}
\label{fig4}
\end{figure}

For $t<0$, let $\Tb^{\text{X}}(t)$ denote the temperature of the reservoir in contact with sample X. We consider protocols in which each reservoir temperature undergoes at most two quenches, at $t=-\tw$ and $t=0$:
\beq
\label{3}
\Tb^\text{X}(t):
\TI^\text{X} \stackrel{t=-\tw}{\longrightarrow}
\TII^\text{X} \stackrel{t=0}{\longrightarrow} \Tc,
\quad \text{X}=\mathrm{A},\mathrm{B}.
\eeq
If $\TII^\text{X}=\TI^\text{X}$ or $\TII^\text{X}=\Tc$, the schedule for sample X reduces to a single-step quench at $t=0$ or $t=-\tw$, respectively. Across the two samples, the protocols involve at most five reservoir temperatures: $\TIA$, $\TIIA$, $\TIB$, $\TIIB$, and $\Tc$. We take the waiting time to be the same for both samples, $\tw^{\text{A}}=\tw^{\text{B}}=\tw$.
Figure~\ref{fig4}(a) depicts schematically this class of protocols.

Our aims are (i) to determine the conditions on the waiting time $\tw$ and the reservoir temperatures $\{\TIA,\TIIA,\TIB,\TIIB,\Tc\}$ for an ME to occur, and (ii)  to identify which protocol maximizes the magnitude of the effect.

\begin{table}%[H]
\caption{Reservoir temperatures in the preparation stages, $t<0$, for five possible  protocols. For all of them, the reservoir in contact with each sample is at temperature $\Tc$ for $t>0$.
The reduced temperature differences $\alAB$, $\beAB$, and $\ww$ are defined by Eqs.~\eqref{12}, \eqref{16a}, and \eqref{23w}, respectively.\label{table1}}
\newcolumntype{C}{>{\centering\arraybackslash}X}
\begin{tabularx}{\textwidth}{lCCCCCC}
\toprule
\textbf{Protocol}	& \boldmath{$\TIA$} 	& \boldmath{$\TIIA$} & \boldmath{$\TIB$} &\boldmath{$\TIIB$}&\boldmath{$\alAB$}&\boldmath{$\beAB$}\\
\midrule
Pontus&$\Tw$&$\Th$&$\Tw$&$\Tc$&$0$&$-1$\\
Two-reservoir&$\Th$&$\Tc$&$\Tc$&$\Th$&$1$&$1$\\
Descartes&$\Th$&$\Tc$&$\Tw$&$\Tw$&$\ww^{-1}-1$&$\ww$\\
WHC&$\Th$&$\Tc$&$\Tw$&$\Th$&$1-\ww$&$1$\\
CWC&$\Th$&$\Tc$&$\Tc$&$\Tw$&$\ww^{-1}$&$\ww$\\
\bottomrule
\end{tabularx}
\end{table}

Table~\ref{table1} and Figs.~\ref{fig4}(b--f) summarize five protocols in this class. The two extreme reservoir temperatures are $\Th$ and $\Tc$. The two-reservoir protocol uses only these temperatures; the other protocols also involve the intermediate temperature $\Tw$. In the Pontus protocol~\cite{NE25,NEDG25}, both samples are initially equilibrated at $\Tw$ for $t<-\tw$. The Descartes protocol~\cite{S26b} uses single-step quenches for both samples: sample A is quenched at $t=-\tw$, and sample B at $t=0$. In the two-reservoir, Descartes, WHC, and CWC protocols, sample A follows the same single-step schedule, $\Th\to\Tc$ at $t=-\tw$; these protocols differ in the preparation of sample B. The acronyms WHC and CWC refer to the successive reservoir temperatures experienced by sample B: $\Tw\to\Th\to\Tc$ and $\Tc\to\Tw\to\Tc$, respectively~\cite{note_26_11_1}.  The WHC and CWC protocols can be viewed as extensions of the two-reservoir protocol: setting $\Tw=\Tc$ in WHC or $\Tw=\Th$ in CWC recovers the two-reservoir schedule.

\section{The model}
\label{sec3}

Newton's law of cooling makes the rate of temperature change proportional to the instantaneous temperature difference between the sample and the bath. Here we study its time-delayed extension,
\beq
\label{eq:delay-law}
\dot{T}(t)=-\left[T(t-\tau)-\Tb(t)\right],
\eeq
where time has been rescaled by the heat-transfer rate.
The bath temperature $\Tb(t)$ may vary with time, and $\tau$ is the dimensionless delay time. We consider $0<\tau<\ee^{-1}\simeq 0.368$. The delayed temperature in Eq.~\eqref{eq:delay-law} makes the relaxation rate depend on the system's earlier state. The equation therefore provides a simple phenomenological model for thermal relaxation with memory~\cite{S24,S25,S26b}.

The delay $\tau$ represents an effective memory time. It can capture, for example, a finite response time of the environment, heat transfer through intermediate degrees of freedom, or memory induced by coarse-graining when faster variables are eliminated. Equation~\eqref{eq:delay-law} is thus a minimal model for examining how such memory modifies thermal relaxation.

For the protocols considered here, the bath is held at the target temperature $\Tc$ for $t>0$, and we require
\beq
\label{42}
\Tc \leq \inf_{t<0}\Tb^\text{X}(t),
\quad \text{X}=\mathrm{A},\mathrm{B}.
\eeq
Consequently, $T_\mathrm{X}(t-\tau)>\Tc$ during the relaxation, and Eq.~\eqref{eq:delay-law} gives $\dot T_\mathrm{X}(t)<0$ for $t>0$. Thus, neither sample overshoots $\Tc$. This monotonic decay does not by itself exclude multiple crossings of $\TA(t)$ and $\TB(t)$. We therefore restrict the analysis to protocols for which the two temperature curves cross at most once.

For a constant bath, the normalized relaxation function is the quasi-exponential function $\EE(t)$, defined for $t\geq 0$ by~\cite{S24}
\bal
\label{eq:quasi-exp}
\EE(t)&=1+\sum_{n=0}^{\lfloor t/\tau \rfloor}
\frac{(n\tau-t)^{n+1}}{(n+1)!},\nn
&=
\begin{cases}
1-t, & 0\leq t\leq \tau,\\[2pt]
1-t+\dfrac{(\tau-t)^2}{2}, & \tau\leq t\leq 2\tau,\\[6pt]
1-t+\dfrac{(\tau-t)^2}{2}
+\dfrac{(2\tau-t)^3}{3!}, & 2\tau\leq t\leq 3\tau,\\
\vdots\ & \vdots
\end{cases}
\eal
where $\lfloor \cdot \rfloor$ denotes the floor function. We set $\EE(t)=1$ for $t\leq 0$, so that
\beq
\label{eq:quasi-exp-derivative}
\dot{\EE}(t)=-\EE(t-\tau).
\eeq

At long times, the leading decay is exponential~\cite{S24,S25}:
\beq
\label{eq:asymptotic-E}
\EE(t)\approx
\frac{\kappa^{-1}}{1-\kappa\tau}\ee^{-\kappa t},
\quad t\gg 1,
\eeq
where $\kappa$ is the root of the equation $\kappa=\ee^{\kappa\tau}$, i.e.,
\beq
\kappa=-\tau^{-1}W_0(-\tau).
\eeq
Here, $W_0$ is the principal branch of the Lambert $W$ function~\cite{CGHJK96}.

\section{Conditions for the Mpemba effect}
\label{sec4}

For the bath schedule in Eq.~\eqref{3}, the method of steps gives the sample temperatures listed in Table~\ref{table2}.
This solution  is a special case of the solution for a generic time-dependent bath protocol derived in the Appendix.

Subtracting the two solutions eliminates the common target temperature and yields, for $t\geq0$,
\beq
\label{TA-TB}
\TA(t)-\TB(t)
=(\aAB+\bAB)\EE(t+\tw)-\bAB\EE(t),\quad t\geq 0,
\eeq
where
\beq
\label{albeAB}
\aAB\equiv\TIA-\TIB,\quad
\bAB\equiv\TIIB-\TIIA.
\eeq

\begin{table}[H]
\caption{Sample temperature $T_\text{X}(t)$ generated by Eq.~\eqref{eq:delay-law} for the bath schedule in Eq.~\eqref{3}.\label{table2}}
\newcolumntype{C}{>{\centering\arraybackslash}X}
\begin{tabularx}{\textwidth}{LL}
\toprule
%\multicolumn{2}{c}{\boldmath{$T_\text{X}(t)$}}\\
&\boldmath{$T_\text{X}(t)$}\\
\midrule
$t\leq -\tw$:&$\TI^\text{X}$\\
$-\tw\leq t\leq 0$:&$\TII^\text{X}+(\TI^\text{X}-\TII^\text{X})\EE(t+\tw)$\\
$t\geq 0$:&$\Tc+(\TII^\text{X}-\Tc)\EE(t)
+(\TI^\text{X}-\TII^\text{X})\EE(t+\tw)$\\
\bottomrule
\end{tabularx}
\end{table}

\begin{figure}
%\begin{adjustwidth}{-\extralength}{0cm}
\centering
\includegraphics[width=0.6\linewidth]{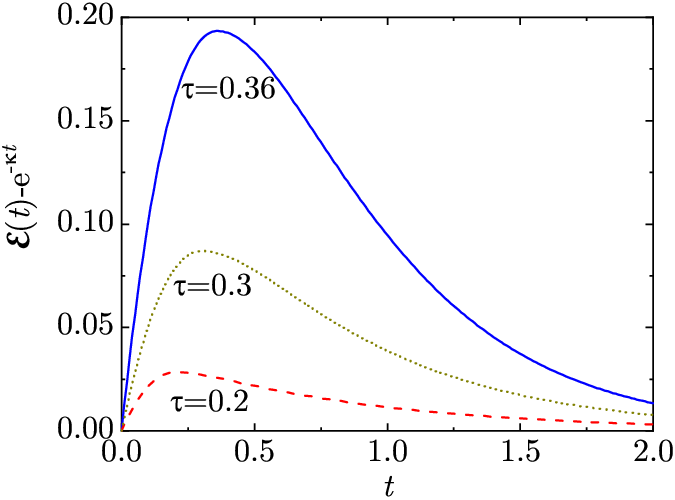}
%\end{adjustwidth}
\vspace{-3pt}
\caption{Dependence of the difference $\EE(t)-\ee^{-\kappa t}$ on $t$ for $\tau=0.2$, $0.3$, and $0.36$, from bottom to top.}
\label{fig1}
\end{figure}

We impose the initial ordering $\TA(0)>\TB(0)$, as in Eq.~\eqref{1}. An ME requires the ordering to reverse at sufficiently long times, $\TA(t)<\TB(t)$, as in Eq.~\eqref{2}. The first condition, evaluated using Eq.~\eqref{TA-TB} at $t=0$, gives
\beq
\aAB+\bAB>\frac{\bAB}{\EE(\tw)}.
\eeq
For the long-time condition, the dominant exponential decay of $\EE$, Eq.~\eqref{eq:asymptotic-E}, gives
\beq
\lim_{t\to\infty}\frac{\EE(t+\tw)}{\EE(t)}
=\ee^{-\kappa\tw}.
\eeq
Thus, the coefficient of the leading long-time contribution to $\TA(t)-\TB(t)$ must be negative:
\beq
(\aAB+\bAB)\ee^{-\kappa\tw}-\bAB<0.
\eeq
Together, these two requirements give the necessary and sufficient conditions
\beq
\label{7}
\frac{\bAB}{\EE(\tw)}
<\aAB+\bAB
<\ee^{\kappa\tw}\bAB.
\eeq
Here and below, $\tw>0$.

The interval in Eq.~\eqref{7} can be nonempty only if
$\bAB<\EE(\tw)\ee^{\kappa\tw}\bAB$. Since
$\EE(t)>\ee^{-\kappa t}$ for $t>0$ (see Fig.~\ref{fig1}), this inequality requires $\bAB>0$. Also, $\EE(\tw)<1$, so the lower bound in Eq.~\eqref{7} then requires $\aAB>0$. Consequently, an ME is possible only if
\beq
\label{8}
\TIA>\TIB,\qquad \TIIA<\TIIB.
\eeq
Both inequalities are taken into account in the schematic schedule of Fig.~\ref{fig4}(a) and hold for the two-reservoir, Descartes, WHC, and CWC  protocols [Table~\ref{table1} and Figs.~\ref{fig4}(c--f)]. The Pontus protocol instead has $\aAB=0$ and $\bAB=\Tc-\Th<0$, and therefore cannot exhibit the effect in this model.

\begin{figure}
%\begin{adjustwidth}{-\extralength}{0cm}
\centering
\includegraphics[width=0.6\linewidth]{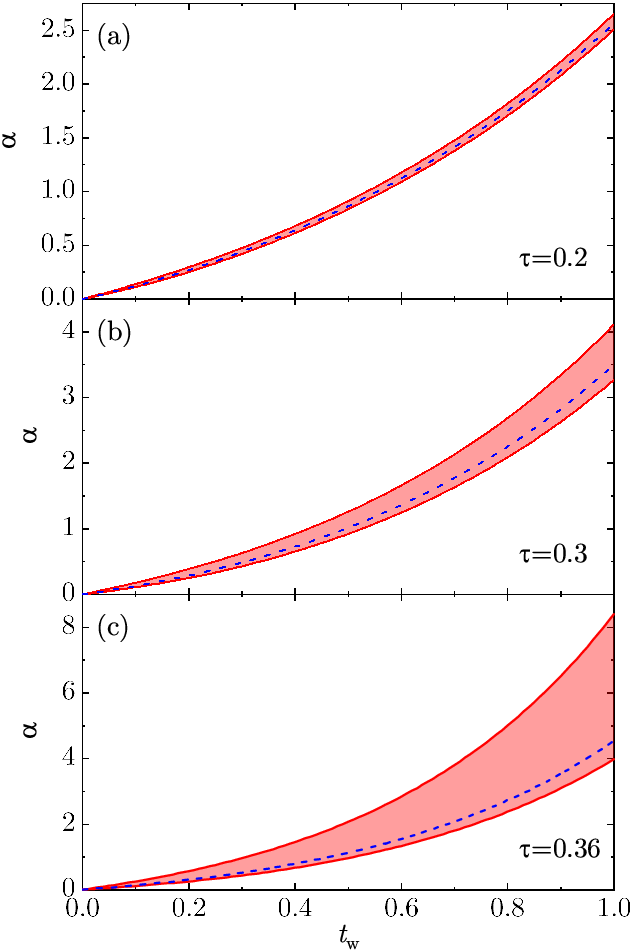}
%\end{adjustwidth}
\vspace{-3pt}
\caption{Lower and upper bounds on $\alAB$ in Eq.~\eqref{9} as functions of $\tw$ for (a) $\tau=0.2$, (b) $\tau=0.3$, and (c) $\tau=0.36$. The shaded region marks the parameter values for which the ME occurs. In each panel, the dashed curves give the balanced optimal value of $\alAB$ for each  $\tw$ (see Sec.~\ref{sec5}).}
\label{fig2}
\end{figure}

It is useful to express Eq.~\eqref{7} in terms of the ratio
\beq
\label{12}
\alAB\equiv\frac{\aAB}{\bAB}
=\frac{\TIA-\TIB}{\TIIB-\TIIA}.
\eeq
Provided Eq.~\eqref{8} holds, the effect occurs precisely when
\beq
\label{9}
\frac{1}{\EE(\tw)}-1
<\alAB
<\ee^{\kappa\tw}-1.
\eeq
Figure~\ref{fig2} shows the allowed region in the $(\tw,\alAB)$ plane for $\tau=0.2$, $0.3$, and $0.36$. The interval between the bounds grows as $\tau$ increases.

To compare the crossover amplitude across protocols, we normalize the sample-temperature difference by the full reservoir-temperature range. We take $\Tc$ to be the lowest reservoir temperature, Eq.~\eqref{42}. Under the necessary conditions in Eq.~\eqref{8}, the highest reservoir temperature is either $\TIA$ or $\TIIB$, so that
\beq
\Th=\max(\TIA,\TIIB).
\eeq
The normalized difference is then
\bal
\label{15}
\Delta(t)\equiv
\frac{\TA(t)-\TB(t)}{\Th-\Tc}
=\beAB D(t),
\eal
with
\begin{subequations}
\label{16}
\beq
\label{16a}
\beAB\equiv\frac{\TIIB-\TIIA}{\Th-\Tc},
\eeq
\beq
\label{16b}
D(t)\equiv(1+\alAB)\EE(t+\tw)-\EE(t).
\eeq
\end{subequations}
Thus, the reservoir temperatures enter $\Delta(t)$ only through the dimensionless combinations $\alAB$ and $\beAB$ (in addition to the delay and waiting times). These two combinations are constrained:
\beq
\alAB\beAB
=\frac{\TIA-\TIB}{\Th-\Tc}\leq 1.
\eeq
Since the reservoir temperatures lie between $\Tc$ and $\Th$, also $\beAB\leq1$. Hence
\beq
\label{22}
0<\beAB\leq\min\!\left(1,\alAB^{-1}\right).
\eeq

The function $D(t)$ has the same form as the normalized difference for the Descartes protocol if its reduced warm temperature
\beq
\label{23w}
\ww\equiv\frac{\Tw-\Tc}{\Th-\Tc}
\eeq
is replaced by $\ww\to (1+\alAB)^{-1}$. Results previously obtained for the Descartes protocol~\cite{S26b} can therefore be adapted to the broader family considered here.

Notice that, in general, the normalized differences of reservoir temperatures, $\alAB$ and $\beAB$, are mutually independent, except for the constraint given by Eq.~\eqref{22}. However, this is not the case of the two-reservoir, Descartes, WHC, and CWC protocols (see Table\ref{table1}), where $\beAB=\alAB=1$, $\beAB=(1+\alAB)^{-1}$, $\beAB=1$, and $\beAB=\alAB^{-1}$, respectively. Moreover, in the two-reservoir, WHC, and CWC protocols one actually has $\beta=\min\!\left(1,\alAB^{-1}\right)$, while $\beta<\min\!\left(1,\alAB^{-1}\right)$ in the Descartes protocol.

\section{Maximizing the Mpemba effect}
\label{sec5}

For fixed $\tau$ and $\tw$, Eq.~\eqref{9} specifies the range of $\alAB$ for which a Mpemba crossover occurs. The crossover becomes weak near either endpoint of this interval. Near the lower endpoint, the initial temperature difference $\TA(0)-\TB(0)$ approaches zero, and the two temperature trajectories cross almost immediately. Near the upper endpoint, the reversal is postponed until both samples are already close to the target temperature $\Tc$.

\begin{figure}
%\begin{adjustwidth}{-\extralength}{0cm}
\centering
\includegraphics[width=0.6\linewidth]{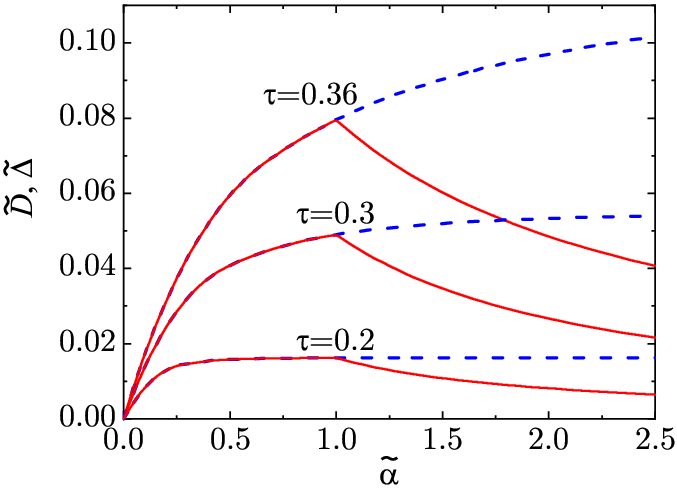}
%\end{adjustwidth}
\vspace{-3pt}
\caption{Parametric plots of $\tD$ (dashed curves) and $\tDelta$ (solid curves) versus $\talAB$, for $\tau=0.2$, $0.3$, and $0.36$, from bottom to top.}
\label{fig3}
\end{figure}

Therefore, we quantify the effect through the extrema of the normalized difference $\Delta(t)=\beAB D(t)$ introduced in Eq.~\eqref{15}:
\beq
\label{24}
\text{Magnitude}=\min\left(\Delta(0),|\Delta_{\min}|\right),
\eeq
where $\Delta_{\min}$ is the minimum of $\Delta(t\geq 0)$.
At fixed $\tau$ and $\tw$, the most balanced crossover is obtained by choosing $\alAB=\talAB$ so that the positive maximum and negative minimum of $D(t)$ have the same absolute value. In this case,
\beq
\tD=D(0)=|D_{\min}|.
\eeq
Substitution of $D(0)=(1+\alAB)\EE(\tw)-1$ [see Eq.~\eqref{16b}] gives
\beq
\label{17}
\talAB=\frac{1+\tD}{\EE(\tw)}-1.
\eeq

The value of $\tD$ for given $\tau$ and $\tw$ can be determined numerically. At fixed $\tau$, both $\talAB$ and $\tD$ increase with $\tw$. An accurate approximation, based on the analysis of the Descartes protocol in Ref.~\cite{S26b}, is
\beq
\label{18}
\tD\simeq\left[\frac{2 - \tau}{1 - \tau - \kappa^{-1}}\frac{1 - \kappa (1 - \kappa \tau)}{1- \ee^{-\kappa \tw}/\EE(\tw)}-1\right]^{-1}.
\eeq
Nevertheless. in what follows we will use the numerical solution.

Figure~\ref{fig3} shows the resulting parametric dependence of $\tD$ and the normalized effect magnitude $\tDelta$ on $\talAB$ for three representative delay times.
Increasing $\tw$ raises $\tD$, but it does not necessarily increase the physical effect magnitude because the prefactor $\beAB$ is constrained by the reservoir-temperature range, Eq.~\eqref{22}. Saturating this constraint gives
\beq
\tDelta=
\begin{cases}
\tD, & \talAB<1,\\[2pt]
\tD/\talAB, & \talAB>1.
\end{cases}
\eeq
As shown in Fig.~\ref{fig3}, at fixed $\tau$ the largest value of $\tDelta$ is attained at $\alAB=\beAB=1$. These values correspond to
\beq
\TIA=\TIIB=\Th,\quad
\TIB=\TIIA=\Tc,
\eeq
that is, to the two-reservoir protocol studied in Ref.~\cite{S24}.

\begin{figure}
%\begin{adjustwidth}{-\extralength}{0cm}
\centering
\includegraphics[width=0.6\linewidth]{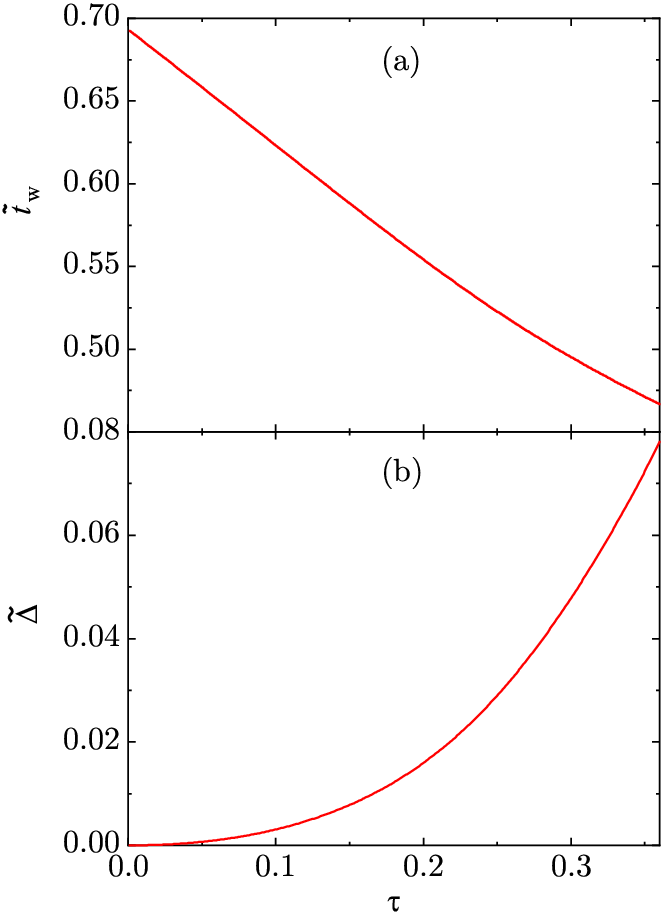}
%\end{adjustwidth}
\vspace{-3pt}
\caption{Optimal waiting time $\ttw$ (a) and corresponding maximum effect magnitude $\tDelta$ (b) versus the delay time $\tau$ for the two-reservoir protocol.}
\label{fig5}
\end{figure}

For that protocol, the optimal waiting time $\ttw$ and the associated maximum magnitude $\tDelta$ vary with the delay time as shown in Fig.~\ref{fig5}. The optimal waiting time exceeds $\tau$ but decreases as $\tau$ increases, whereas the maximum effect magnitude grows with $\tau$.

Table~\ref{table3} compares the two-reservoir optimum with the Descartes protocol and two three-reservoir examples. In each row, $\tw$ is optimized for the specified values of $\alAB$ and the largest allowed $\beAB$. The examples show how the effect magnitude decreases when $\alAB$ differs from its globally optimal value of unity.

\begin{table}[H]
\caption{Optimal waiting time $\ttw$ and corresponding maximum effect magnitude $\tDelta$ for the two-reservoir protocol, the Descartes protocol, and two three-reservoir examples with $\alAB\ne1$. Except in the Descartes protocol, one has $\beAB=\min(1,\alAB^{-1})$. The delay time is $\tau=0.36$ in all the cases.\label{table3}}
\newcolumntype{C}{>{\centering\arraybackslash}X}
\begin{tabularx}{\textwidth}{CCCCC}
\toprule
\textbf{Protocol}	&\boldmath{$\alAB$}&\boldmath{$\beAB$}&\boldmath{$\ttw$}&\boldmath{$\tDelta$}\\
\midrule
Two-reservoir&$1$&$1$&$0.466$&$0.080$\\
Descartes&$0.668$&$0.600$&$0.36$&$0.040$\\
WHC&${4}/{5}$&$1$&$0.405$& $0.073$\\
CWC&${6}/{5}$&${5}/{6}$&$0.520$&$0.070$\\
\bottomrule
\end{tabularx}
\end{table}

\section{Protocols beyond the synchronous two-reservoir case}
\label{sec6}
Having established that the synchronous two-reservoir protocol maximizes the effect magnitude among protocols with at most two quenches and a common waiting time, we now test the robustness of this optimum by relaxing these restrictions separately, beginning with asynchronous quenches and then allowing an additional quench for sample B.

\subsection{Two-reservoir protocol with asynchronous quenches}

\begin{figure}
%\begin{adjustwidth}{-\extralength}{0cm}
\centering
\includegraphics[width=0.6\linewidth]{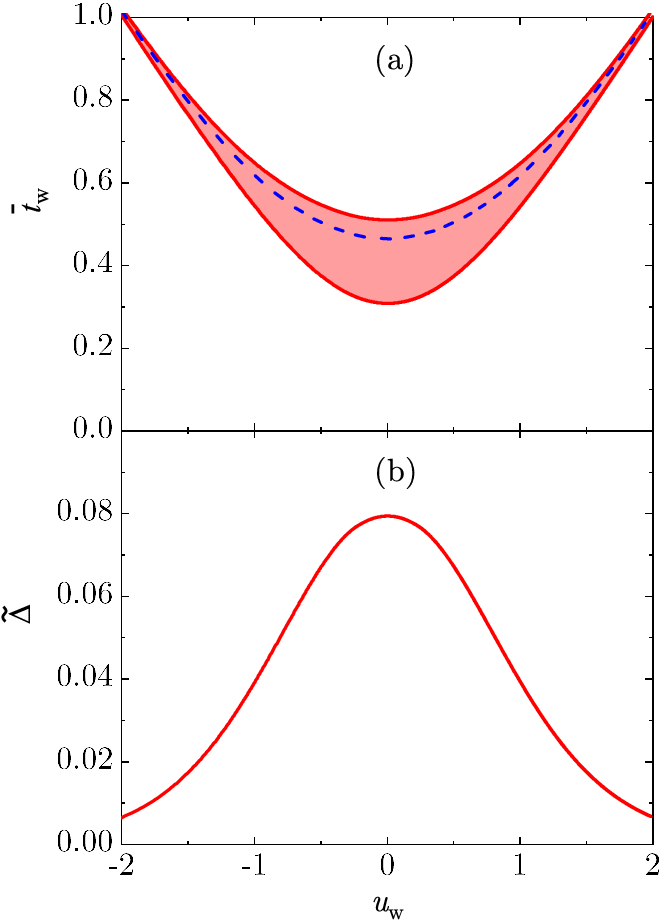}
%\end{adjustwidth}
\vspace{-3pt}
\caption{(a) Region in the $(\uw,\bartw)$ plane [see Eq.~\eqref{31}] where the ME occurs for the two-reservoir protocol with asynchronous quenches at $\tau=0.36$. The dashed curve gives the balanced optimal mean waiting time $\ttw$ for each separation $\uw$. (b) Corresponding balanced effect magnitude $\tDelta$.}
\label{fig6}
\end{figure}

We now allow the single quench of sample A and the first quench of sample B to occur at different times, $t=-\twA$ and $t=-\twB$, respectively. For $t\geq0$, the normalized temperature difference is
\beq
\Delta(t)=\EE(t+\twA)+\EE(t+\twB)-\EE(t).
\eeq
This expression is symmetric under exchange of $\twA$ and $\twB$

The initial ordering $\Delta(0)>0$ requires $\EE(\twA)+\EE(\twB)>1$. At long times, the leading exponential term in $\EE(t)$, Eq.~\eqref{eq:asymptotic-E}, determines the sign of $\Delta(t)$.
Thus, $\Delta(t)$ is eventually negative precisely when
$\ee^{-\kappa\twA}+\ee^{-\kappa\twB}<1$. The conditions for a crossover are therefore
\beq
\label{30}
\ee^{-\kappa\twA}+\ee^{-\kappa\twB}
<1<
\EE(\twA)+\EE(\twB).
\eeq
For $\twA=\twB=\tw$, these reduce to Eq.~\eqref{9} with $\alAB=1$.

To describe the schedules in terms of their average and their separation, we introduce
\beq
\label{31}
\bartw\equiv\frac{\twA+\twB}{2},
\quad
\uw\equiv\twA-\twB.
\eeq
Figure~\ref{fig6}(a) shows the region satisfying Eq.~\eqref{30} for $\tau=0.36$.
Since $\Delta(t)$ is invariant under the exchange $\twA\leftrightarrow\twB$, the results are symmetric under $\uw\to-\uw$. We therefore display both signs of $\uw$, which correspond to the two possible orderings of the waiting times.
The admissible interval of mean waiting times narrows as $|\uw|$ increases and closes at the boundary of the region.

Let $\tx$ denote the crossover time. The difference function obeys $\dot{\Delta}(t)=-\Delta(t-\tau)$ [see Eq.~\eqref{eq:quasi-exp-derivative}].
Since $\Delta$ is positive before its single crossover at $t_x$ and negative afterward, this relation makes $\Delta(t)$ decrease until $t=t_x+\tau$ and increase thereafter; hence its minimum occurs at $t_x+\tau$: $\Delta_{\min}=\Delta(\tx+\tau)$.
For each fixed separation $\uw$, we determine the balanced optimal mean waiting time $\ttw$ by imposing $\Delta(0)=|\Delta_{\min}|$, the corresponding effect magnitude being $\tDelta=\Delta(0)$. The dashed line in Fig.~\ref{fig6}(a) shows $\ttw$ as a function of $\uw$, and Fig.~\ref{fig6}(b) gives the resulting $\tDelta$ at $\tau=0.36$.  The largest magnitude occurs for $\uw=0$, indicating that asynchronous quenches weaken the effect within this two-reservoir family.
Although Fig.~\ref{fig6} shows the case $\tau=0.36$, repeating the calculation for other values of the delay time ($\tau=0.05, 0.10, 0.15, 0.20, 0.25, 0.30$) confirms the same conclusion: the maximum effect magnitude occurs for synchronous quenches, $\uw=0$.

\subsection{Three-quench protocol for sample B}

We next test whether the synchronous two-reservoir protocol remains optimal when sample B is allowed one additional quench, while sample A retains the schedule $\Th\to\Tc$ at $t=-\tw$.
The schedule for sample B will be of the form
\beq
\label{32}
\Tb^{\text{B}}(t):
\Tc \stackrel{t=-\twI}{\longrightarrow}
\TIIB \stackrel{t=-\twII}{\longrightarrow} \TIIIB\stackrel{t=0}{\longrightarrow}
\Tc.
\eeq
We take $0\leq\twII\leq\twI$, so the switches at $-\twI$ and $-\twII$ occur in the stated order.
For $t\geq 0$, the solution to Eq.~\eqref{eq:delay-law} is (see the Appendix)
\beq
\label{33}
\TB(t)=\Tc+(\TIIIB-\Tc)\EE(t)+(\TIIB-\TIIIB)\EE(t+\twII)+(\Tc-\TIIB)\EE(t+\twI),\quad t\geq 0.
\eeq

One of the two quenches at $t=-\twI$ and $t=-\twII$ is assumed to be simultaneous to the single quench of sample A at $t=-\tw$, and thus either $(\twI,\twII)=(\tw,\tw')$ or $(\twI,\twII)=(\tw',\tw)$. Finally, $\Tw=\min(\TIIB,\TIIIB)$ and $\Th=\max(\TIIB,\TIIIB)$ are the lower and higher intermediate reservoir temperatures, respectively.
Since all reservoir temperatures are restricted to the interval $[\Tc,\Th]$, these definitions imply $\Tc\leq\Tw\leq\Th$.
The four possible schedules for sample B are listed in Table~\ref{table4}.
They reduce to the synchronous two-quench case in the following limits: (B1) $\Tw=\Th$ or $\tw'=\tw$, (B2) $\Tw=\Tc$ or $\tw'=\tw$, (B3) $\Tw=\Th$ or $\tw'=0$, (B4) $\Tw=\Th$ and $\tw'=\tw$.

\begin{table}[H]
\caption{Three-quench schedules for sample B [see Eq.~\eqref{32}]. Sample A follows $\Th\to\Tc$ at $t=-\tw$ in all cases.\label{table4}}
\newcolumntype{C}{>{\centering\arraybackslash}X}
%\begin{tabularx}{\textwidth}{CCCCCC}
\begin{tabularx}{\textwidth}{cccccc}
\toprule
\textbf{Schedule}	& \boldmath{$\TIIB$} 	& \boldmath{$\TIIIB$} & \boldmath{$\twI$} &\boldmath{$\twII$}&\boldmath{$\Delta(t)$}\\
\midrule
\BI   & $\Tw$&$\Th$&$\tw$&$\tw'$&$(1+\ww)\EE(t+\tw)+(1-\ww)\EE(t+\tw')-\EE(t),\quad \tw'\leq\tw$\\
\BII  & $\Tw$&$\Th$ & $\tw'$&$\tw$&$(2-\ww)\EE(t+\tw)+\ww\EE(t+\tw')-\EE(t),\quad \tw'\geq\tw$\\
\BIII & $\Th$&$\Tw$&$\tw$&$\tw'$&$2\EE(t+\tw)-(1-\ww)\EE(t+\tw')-\ww\EE(t),\quad \tw'\leq\tw$\\
\BIV  & $\Th$&$\Tw$&$\tw'$&$\tw$&$\ww\EE(t+\tw)+\EE(t+\tw')-\ww\EE(t),\quad \tw'\geq\tw$\\
\bottomrule
\end{tabularx}
\end{table}

\subsubsection{Schedules B1 and B2}

\begin{figure}
%\begin{adjustwidth}{-\extralength}{0cm}
\centering
\includegraphics[width=0.6\linewidth]{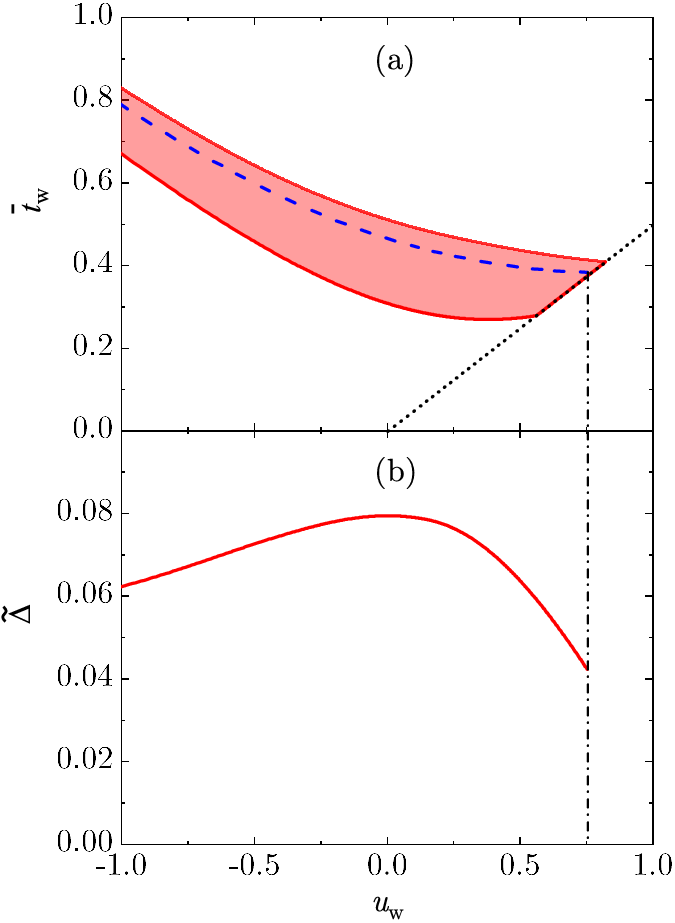}
%\end{adjustwidth}
\vspace{-3pt}
\caption{(a) Region in the $(\uw,\bartw)$ plane [see Eq.~\eqref{35}] where the ME occurs for the  protocols B1 and B2  at $\ww_0=0.4$ and $\tau=0.36$. The dashed curve gives the balanced optimal mean waiting time $\ttw$ for each separation $\uw$. (b) Corresponding balanced effect magnitude $\tDelta$. For B1, $\uw=\tw-\tw'\geq0$, whereas for B2, $\uw\leq0$; the two branches therefore represent the two possible temporal orderings of the pre-zero quenches.}
\label{fig7}
\end{figure}

In the schedules B1 and B2 the difference function has the common form
\beq
\Delta(t)=(1+\ww_0)\EE(t+\tw)+(1-\ww_0)\EE(t+\tw')-\EE(t),
\eeq
with
\bal
\text{B1: }&\ww_0=\ww,\quad 0\leq \tw'\leq\tw,\nn
\text{B2: }&\ww_0=1-\ww,\quad 0\leq \tw\leq\tw'.
\eal

Given $\tau$, $\tw$, and $\ww_0$, the  secondary waiting time $\tw'$ must satisfy the conditions
\beq
\label{34}
(1+\ww_0)\ee^{-\kappa\tw}+(1-\ww_0)\ee^{-\kappa\tw'}
<1<
(1+\ww_0)\EE(\tw)+(1-\ww_0)\EE(\tw')
\eeq
for the existence of ME.

As an illustration, Fig.~\ref{fig7}(a) shows the region described by Eq.~\eqref{34} on the plane $(\bartw,\uw)$ for $\ww_0=0.4$ and $\tau=0.36$, where now
\beq
\label{35}
\bartw\equiv\frac{\tw+\tw'}{2},
\quad
\uw\equiv\tw-\tw'.
\eeq
The shaded region is cut at the right by the line $\bartw=\uw/2$, which corresponds to $\tw'=0$. The dashed line inside the shaded region in Fig.~\ref{fig7}(a) represents the locus $\Delta(0)=|\Delta_{\min}|=\tDelta$. The corresponding dependence of $\tDelta$ on $\uw$ is shown in Fig.~\ref{fig7}(b). We observe that the largest magnitude occurs for $\uw=0$, which corresponds to the synchronous two-reservoir protocol.

\subsubsection{Schedule B3}

\begin{figure}
%\begin{adjustwidth}{-\extralength}{0cm}
\centering
\includegraphics[width=0.6\linewidth]{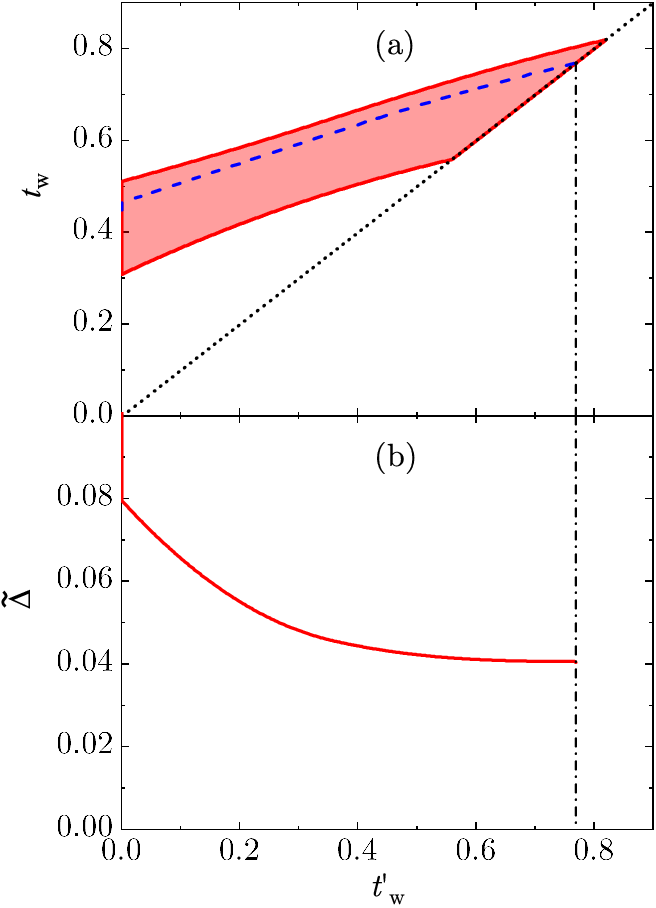}
%\end{adjustwidth}
\vspace{-3pt}
\caption{(a) Region in the $(\tw',\tw)$ plane  where the ME occurs for the  protocol B3  at $\ww=0.4$ and $\tau=0.36$. The dashed curve gives the balanced optimal waiting time $\ttw$ for each  $\tw'$. (b) Corresponding balanced effect magnitude $\tDelta$.}
\label{fig8}
\end{figure}

In this case, the ME requires
\beq
2\ee^{-\kappa\tw}-(1-\ww)\ee^{-\kappa\tw'}<\ww<2\EE(\tw)-(1-\ww)\EE(\tw'),
\eeq
with $0\leq\tw'\leq\tw$.

An analysis similar to the previous ones yields Fig.~\ref{fig8} for $\ww=0.4$ and $\tau=0.36$. Again, the maximum effect occurs for the synchronous two-reservoir protocol ($\tw'=0$).

\subsubsection{Schedule B4}

\begin{figure}
%\begin{adjustwidth}{-\extralength}{0cm}
\centering
\includegraphics[width=0.6\linewidth]{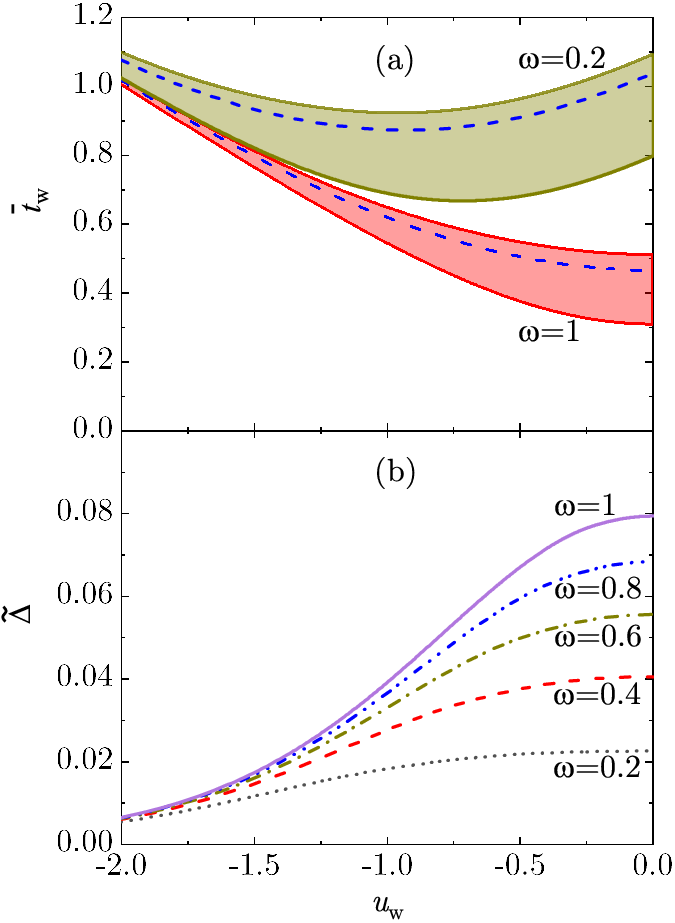}
%\end{adjustwidth}
\vspace{-3pt}
\caption{(a) Region in the $(\uw,\bartw)$ plane [see Eq.~\eqref{35}] where the ME occurs for the  protocol B4  at $\ww=0.2$ and $\ww=1$ with $\tau=0.36$. The dashed curves give the balanced optimal mean waiting time $\ttw$ for each separation $\uw$. (b) Corresponding balanced effect magnitude $\tDelta$ for $\ww=0.2$, $0.4$, $0.6$, $0.8$, and $1$.}
\label{fig9}
\end{figure}

In the case of this fourth schedule, the ME condition is
\beq
\ww\ee^{-\kappa\tw}+\ee^{-\kappa\tw'}<\ww<\ww\EE(\tw)+\EE(\tw'),
\eeq
with $\tw'\geq\tw\geq 0$.
This region is shown in Fig.~\ref{fig9}(a) in the representation $(\uw,\bartw)$ for $\ww=0.2$ and $\ww=1$ with $\tau=0.36$. Figure~\ref{fig9}(b) presents the balanced magnitude of the effect, $\tDelta$, for $\ww=0.2$, $0.4$, $0.6$, $0.8$, and $1$. For any value of $\ww$, the maximum occurs at $\uw=0$, i.e., $\tw'=\tw$, in which case the B4 schedule reduces to the CWC one [see Table~\ref{table1} and Fig.~\ref{fig4}(f)]. As expected, the absolute maximum corresponds to $\ww=1$, that is, the two-reservoir protocol.

\subsection{A conjecture}

By repeating the numerical calculations for the schedules B1--B4 with other combinations of $\ww$ and $\tau$, namely $\ww=0,0.2,0.4,0.6,0.8,1$ and $\tau=0.05, 0.10, 0.15, 0.20, 0.25, 0.30, 0.36$,  it is found again that the largest effect magnitude occurs when the schedules reduce to the synchronous two-reservoir protocol. This provides numerical support for the ansatz that the synchronous two-reservoir protocol is optimal beyond the semi-analytically established class considered in Secs.~\ref{sec2}--\ref{sec5}. The numerical searches cover only the parameter values and ranges specified above and do not prove the conjecture for arbitrary protocols.

To sum up, we have proved semi-analytically that, among protocols with at most two quenches and a common waiting time, the synchronous two-reservoir protocol yields the maximum effect magnitude. Numerical explorations of the asynchronous two-reservoir protocol and the four three-quench protocols B1--B4 show the same maximum within the parameter ranges examined. Motivated by these results, we conjecture that the synchronous two-reservoir protocol remains optimal for the broader class of preparation histories defined below.

Consider a generic pair of preparation protocols, $\Tb^\text{X}(t<0)$,
with $\text{X}=\mathrm{A},\mathrm{B}$, followed by a quench to the common
target temperature $\Tc$ at $t=0$:
\beq
\label{41}
\Tb^\text{X}(t<0) \stackrel{t=0}{\longrightarrow} \Tc.
\eeq
The bath histories may include continuous temperature variations, discrete
quenches, or combinations of both. To specify the initial condition for the
delay equation, we assume that each sample is equilibrated with its initial
reservoir before a finite preparation interval, during which its bath
temperature may vary. The histories are bounded and satisfy Eq.~\eqref{42}.
The temperature evolution for such general bath histories follows from the solution derived in the Appendix  [see Eq.~\eqref{A5}].

For each comparison, the two protocols share the same target temperature
$\Tc$ and the same upper temperature $\Th$, defined by
\beq
\label{43}
\Th = \sup_{\substack{t<0}}
\Tb^\text{X}(t),
\quad  \text{X}=\mathrm{A},\mathrm{B}.
\eeq
Thus, all bath temperatures lie in the common interval $[\Tc,\Th]$.
Restricting attention to protocols for which the sample temperatures cross
at most once, we define the normalized temperature difference $\Delta(t)$
as in Eq.~\eqref{15}. If the protocol produces an ME, its magnitude
is defined by Eq.~\eqref{24}.

Under those conditions our conjecture states that the protocol that maximizes the magnitude of the ME is the synchronous two-reservoir protocol with $\tw=\ttw$.
In particular, for $\tau=0.36$, the optimized synchronous two-reservoir protocol gives a maximum temperature separation $\max_{t\geq 0}|T_\mathrm{B}(t)-T_\mathrm{A}(t)|\simeq 0.08(\Th-\Tc)$; thus, the peak separation is approximately $8\%$ of the fixed temperature span.

\section{Conclusions}
\label{sec7}

The ME has been investigated within the time-delayed Newton
cooling law for samples prepared by different bath protocols and then
exposed to a common target bath. For the class of protocols involving at
most two quenches and a common waiting time, necessary and sufficient
conditions for the effect have been obtained in terms of the waiting time
and the reservoir temperatures. In particular, the effect requires the
initial reservoir temperatures and the intermediate reservoir temperatures
to be ordered in opposite ways for the two samples. The Pontus protocol
does not satisfy these conditions and therefore cannot produce an ME within this model.

The effect magnitude was then optimized within this protocol class, using
the balanced condition $\Delta(0)=|\Delta_{\min}|$. The synchronous
two-reservoir protocol was shown to yield the largest magnitude. This
result identifies the optimal protocol within the stated class, rather
than among all possible preparations.

The scope of this result was tested numerically by relaxing the
synchrony and quench-number restrictions. For the asynchronous
two-reservoir protocol and the four three-quench protocols considered,
the largest effect magnitude found occurs when the schedules reduce to
the synchronous two-reservoir protocol. These numerical results support
the special role of the synchronous protocol for the parameter ranges
investigated, but do not establish its optimality over all possible
bath histories.

Motivated by the semi-analytical result and these numerical explorations, it is
conjectured that, for a fixed delay time and common temperature interval
$[\Tc,\Th]$, the synchronous two-reservoir protocol maximizes the effect
magnitude over the broader class of admissible preparation histories.
Here, the bath temperature may vary continuously, through discrete
quenches, or through a combination of both. Establishing or disproving
this conjecture remains an open problem.

The conclusions are drawn for protocols whose bath temperatures do not
fall below $\Tc$, so that both samples cool monotonically toward the
target temperature. The analysis is further restricted to cases in which
the sample temperatures cross at most once. Extending the results to
multiple-crossing scenarios, or proving the conjecture for general bath
histories, would clarify the limits of the optimality result.

\vspace{6pt}
%\authorcontributions{Conceptualization, A.S.; methodology, D.P. and A.S.; software, A.S.; validation, D.P. and A.S.; formal analysis, D.P. and A.S.; investigation, D.P. and A.S.;  writing---original draft preparation, A.S.; writing---review and editing, D.P. and A.S.; visualization, A.S.; supervision, A.S.; funding acquisition, A.S. All authors have read and agreed to the published version of the~manuscript.}
\authorcontributions{The author is the sole contributor to this work.}

\funding{I acknowledge financial support from Grant No.~PID2024-156352NB-I00 funded by MCIU/\allowbreak AEI/\allowbreak 10.13039/\allowbreak 501100011033 and by ERDF/EU, and from Grant No.~GR24022 funded by the Junta de Extremadura (Spain).}

\institutionalreview{{~}Not applicable.
}

\dataavailability{The raw data supporting the conclusions of this article will be made available by the authors on request.}

\conflictsofinterest{The author declares no conflicts of~interest.}

\abbreviations{Abbreviations}{
The following abbreviations are used in this manuscript:\\

\noindent
\begin{tabular}{@{}ll}
CWC& Cold--warm--cold\\
ME& Mpemba effect\\
WHC& Warm--hot--cold\\
\end{tabular}
}

\appendixtitles{yes}
\appendixstart
\appendix
\setcounter{equation}{0}
\setcounter{figure}{0}
\section*{\appendixname. Solution for a generic protocol}

Consider a sample whose temperature is $\TI$ for $t\leq0$. For $t>0$,
it is coupled to a bath with prescribed temperature $\Tb(t)$. Taking the
Laplace transform of Eq.~\eqref{eq:delay-law} gives~\cite{S24}
\beq
\label{A1}
\wT(s)=\TI\wEE(s)-\wTb(s)\left[s\wEE(s)-1\right],
\eeq
where
\beq
\label{A2}
\wEE(s)=\frac{1}{s}
-\frac{1}{s(s+\ee^{-s\tau})}.
\eeq
Inverting the transform yields
\bal
\label{A3}
T(t)=&\TI\EE(t)-\int_0^t dt_1\,\Tb(t_1)\dot{\EE}(t-t_1)
%\nn=&\left[\TI-\Tb(0^+)\right]\EE(t)+\Tb(t)-\int_0^t dt_1\,\dot{T}_{\text{b}}(t_1)\EE(t-t_1)
.
\eal

Suppose that the bath reaches the target temperature at a finite time
$\tw$, so that $\Tb(t)=\Tc$ for $t>\tw$. Write
$\Tb(t)=\Tc+\delta\Tb(t)$, where $\delta\Tb(t)=0$ for $t>\tw$.
Using Eq.~\eqref{eq:quasi-exp-derivative}, Eq.~\eqref{A3} becomes
\beq
\label{A4}
T(t)=\Tc+(\TI-\Tc)\EE(t)
+\int_0^{\min(\tw,t)} dt_1\,
\delta\Tb(t_1)\EE(t-t_1-\tau),
\quad t\geq0.
\eeq

\begin{figure}
%\begin{adjustwidth}{-\extralength}{0cm}
\centering
\includegraphics[width=0.6\linewidth]{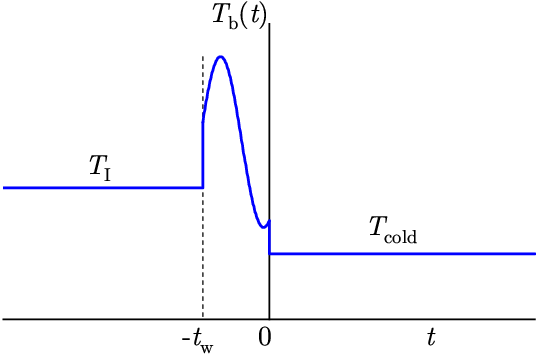}
%\end{adjustwidth}
\vspace{-3pt}
\caption{Schematic bath-temperature history for a generic preparation
protocol. The sample is initially equilibrated with the reservoir at
$\TI$. During the preparation interval $-\tw<t<0$, the bath temperature
may vary continuously, as illustrated, while remaining no lower than
$\Tc$. At $t=0$, the bath is quenched to the target temperature $\Tc$,
which is then held fixed. More general protocols may also include
discrete quenches within the interval $-\tw<t<0$.}
\label{figA1}
\end{figure}

We now shift the time origin to the instant when the bath reaches $\Tc$,
using $t\mapsto t-\tw$.
A schematic example of a preparation protocol in this time coordinate is
shown in Fig.~\ref{figA1}: the bath temperature may vary during
$-\tw<t<0$ before being quenched to $\Tc$ at $t=0$.
To simplify notation, we continue to denote the
temperature and bath histories in the shifted time coordinate by $T(t)$
and $\Tb(t)$, respectively. Thus, the bath deviation vanishes for $t>0$,
and Eq.~\eqref{A4} becomes
\beq
\label{A5}
T(t)=\Tc+(\TI-\Tc)\EE(t+\tw)
+\int_{-\tw}^{\min(0,t)}dt_1\,
\delta\Tb(t_1)\EE(t-t_1-\tau),
\quad t\geq-\tw.
\eeq

For the double-quench protocol, the bath is at $\TI$ for $t<-\tw$,
at $\TII$ for $-\tw<t<0$, and at $\Tc$ for $t>0$. Thus, in
Eq.~\eqref{A5}, $\delta\Tb(t)=\TII-\Tc$ for $-\tw<t<0$. Substitution into
Eq.~\eqref{A5} reproduces the corresponding entries of Table~\ref{table2}.
Likewise, substituting the bath history for the protocol considered in Eq.~\eqref{32} into Eq.~\eqref{A5} recovers Eq.~\eqref{33}.

\begin{adjustwidth}{-\extralength}{0cm}

\reftitle{References}

% Please provide either the correct journal abbreviation (e.g., according to the “List of Title Word Abbreviations” http://www.issn.org/services/online-services/access-to-the-ltwa/) or the full name of the journal.

%\externalbibliography{yes}
%\bibliography{C:/AA_D/Dropbox/Mis_Dropcumentos/bib_files/Granular}

\begin{thebibliography}{999}
\bibitem[not()]{note_26_11_1}
With the same notation, the two-reservoir and Descartes protocols could be
  denoted CHC and WCC, respectively.
\end{thebibliography}

\PublishersNote{}
\end{adjustwidth}

\end{document}